# Enhanced Superconductivity in 2H-TaS$_2$ Devices Through in-situ Molecular Intercalation

*Jose M. Pereira[1]‡, Daniel Tezze[1]‡, Beatriz Martín-García[1,2], Fèlix Casanova[1,2], Maider Ormaza[3], Luis E. Hueso[1,2] and Marco Gobbi[2,4]\*.*

[1] CIC nanoGUNE BRTA, 20018 Donostia-San Sebastián, Spain

[2] IKERBASQUE, Basque Foundation for Science, 48013 Bilbao, Spain

[3] Departamento de Polímeros y Materiales Avanzados: Física, Química y Tecnología (UPV-EHU), 20018 San Sebastián, Spain

[4] Centro de Física de Materiales (CSIC-UPV-EHU) and Materials Physics Center (MPC), 20018 San Sebastián, Spain.

\* Corresponding Author

ABSTRACT:

The intercalation of guest species into the gap of van der Waals materials often leads to the emergence of intriguing phenomena, such as superconductivity. While intercalation-induced superconductivity has been reported in several bulk crystals, reaching a zero-resistance state in flakes remains challenging. Here, we show a simple method for enhancing the superconducting transition in tens-of-nm thick 2H-TaS$_2$ crystals contacted by gold electrodes through in-situ intercalation. Our approach enables measuring the electrical characteristics of the same flake before and after intercalation, permitting us to precisely identify the effect of the guest species on the TaS$_2$ transport properties. We find that the intercalation of amylamine molecules into TaS$_2$ flakes causes a suppression of the charge density wave and an increase in the superconducting transition, with an onset temperature above 3 K. Additionally, we show that a fully developed zero-resistance state can be achieved in flakes by engineering the conditions of the chemical intercalation. Our findings pave the way for the integration of chemically tailored intercalation compounds in scalable quantum technologies.

1. **Introduction**

Intercalation, the insertion of foreign guest species between layers of van der Waals host materials, offers a versatile approach for tailoring their physical properties.[1–4] While this approach has been investigated for decades,[5,6] it has garnered a renewed interest within the domain of the 2D materials research. In particular, recent studies focus on intercalating organic compounds into tens-of-nm thick flakes.[7–10] While molecular intercalation offers tailored chemical tunability[9–11], micro flakes are ideal for integration in micro and nanodevices.

Significantly, intercalation may lead to the emergence[5,6,12] or the enhancement[13] of superconductivity. Intercalation-induced superconductivity has been under study in layered intercalated compounds since the 1960s[5,6] and remains a subject of ongoing investigation.[14–16] Nevertheless, most studies regarding superconductivity in intercalated compounds have focused on bulk crystals.[16–20] This approach struggles to establish a clear path towards technological applications, as achieving superconductivity in exfoliated flakes remains challenging. In a previous study, we illustrated that bulk $MoS_2$ becomes superconductive at low temperatures upon electrochemical intercalation with tetraethylammonium cations.[12] However, no zero-resistance state was found in tens-of-nm thick flakes exfoliated from the intercalated crystals.[12] Therefore, the demonstration of a zero-resistance state in flakes remains a critical yet underexplored step toward utilizing intercalated compounds in functional quantum devices.

2H-$TaS_2$ is an ideal candidate for exploring intercalation-induced superconductivity. It possesses the ability to accommodate diverse guest species, both inorganic[21] and organic[13], and exhibits remarkable resilience to chemical attacks from ambient humidity and oxygen. In its pristine state, 2H-$TaS_2$ exhibits a bulk critical temperature ($T_c$) of approximately 0.8 K[22], enhanced when thinned down[23,24] or intercalated[13,25–31], and enters a charge density wave (CDW) state at around 75 K.[32]

However, even for this material, the studies of intercalation-induced superconductivity have mostly concerned bulk samples.[25–31]

Here, we successfully intercalate amylamine (AM) molecules in tens-of-nm thick $TaS_2$ flakes contacted by metallic electrodes. By comparing the electrical characteristics of pristine and AM-intercalated $2H$-$TaS_2$, we show that molecular intercalation results in an increase of the superconductivity onset to above 3 K. Moreover, we show that a fully developed zero-resistance state can be achieved by engineering the conditions of the chemical intercalation. Our electrical characterization demonstrates that AM-intercalated $TaS_2$ flakes display superconducting characteristics similar to $TaS_2$ monolayers, and unlike the monolayers, they can be exposed to the atmosphere without significantly altering their electrical properties. Our findings offer a potential straightforward route for the application of chemically tailored intercalation compounds in scalable quantum devices.

2. **Experimental Section**

**Chemical intercalation of 2H $TaS_2$**

$2H$-$TaS_2$ crystals were purchased from HQ graphene. Acetonitrile (anhydrous <0.001% $H_2O$, purity 99.8%) and Amylamine (AM – purity >99%) were purchased from Sigma-Aldrich. Solutions are prepared by mixing one part of AM with 2 parts of ACN in volume.

We conduct the intercalation of amylamine (AM) into $TaS_2$ flakes inside a glass vial sealed with a cap. The entire process takes place under atmospheric conditions (in air and at room temperature) and therefore there is an expected amount of water introduced by the solvent and the environment. Different intercalation times were tested, as mentioned in the main text. After removing the substrates from the solution, we rinse them three times with ACN.

**Materials characterization**

XRD measurements were carried out using an Empyrean diffractometer (PANalytical) on bulk crystal and on exfoliated flakes supported on a Si/SiO$_2$ substrate. A copper cathode is used as X-ray source. Both the wavelengths K$\alpha_1$ (1.5406 Å) and K$\alpha_2$ (1.5443 Å) are employed to maximize the intensity of the diffracted beam.

Micro-Raman spectroscopy: Room temperature micro-Raman measurements were carried out in a Renishaw® in Via Qontor instrument equipped with a 100× objective using a 532 nm laser as the excitation source (diffraction grating 1800 l mm$^{-1}$) and an incident power < 1 mW to avoid damage of the sample during the measurements. A linear background was subtracted.

Fabrication of exfoliated samples and device preparation: 2H-TaS$_2$ Flakes were exfoliated from bulk crystals using the scotch tape (Nitto® SPV224P) technique and transferred onto a Si/SiO$_2$ (300nm) substrate to obtain the x-ray diffraction patterns. Homogeneous flakes with a typical thickness of 10 nm - 20 nm were chosen based on the optical contrast. In the case of device preparation, the process takes place using the dry polymer (polydimethylsiloxane, PDMS) technique to transfer the desired flake on top of the prepatterned Ti/Au (5 nm / 15 nm) electrodes using a delamination-stamping system.

Video acquisition: Videos of the intercalation progress were recorded using an optical microscope equipped with a stage operated by a micromanipulator. We prefocus on the desired flake, add a 100-µl droplet of the chosen intercalating medium, and then refocus to obtain the imaging of the flake immersed in the solution. The total intercalation time was typically set to 30 minutes (see main text); the videos included in the Supporting information have been sped up 24 times.

Electrical transport measurements: The electrical measurements were recorded by cooling the devices down to 1.9 K using a Physical Property Measurement System (PPMS, Quantum Design) while measuring the longitudinal resistance as a function of temperature. The resistance was measured using a Keithley 6221 as current-source, and a Keithley 2182 as a nanovoltmeter in delta mode.

## 3. Results and discussion

Prior to the electrical characterization, we investigate the intercalation process for 2H TaS$_2$ flakes exfoliated on a Si/SiO$_2$ substrate utilizing X-ray diffractometry. This technique provides insights into the success of intercalation and the structural features of the intercalated hybrid system. A schematic of our approach is shown in **Figure 1a**. We first use micromechanical exfoliation to obtain 2H-TaS$_2$ flakes and transfer them on the surface of 10 mm × 10 mm Si (300 nm)/SiO$_2$ substrates. Then, we submerge the substrates covered with flakes in a solution containing AM for a set amount of time (Figure 1a). The intercalation of AM molecules occurs spontaneously at room temperature; to stop it, we remove the TaS$_2$ from the solution and rinse it carefully with acetonitrile (ACN). This simple, reliable, and scalable methodology is enabled by 2H TaS$_2$, which is particularly well-suited for chemical intercalation. The primary drive of this process is the charge transfer between guest molecules and host materials, which requires a suitable alignment of the highest occupied molecular orbital (HOMO) of the guest and the work function of the host.[33] The TaS$_2$ work function, approximately 5.6 eV,[34] is relatively close to the HOMO of organic amines (approximately 6 eV) [33], permitting a facile chemical intercalation. Unlike other transition metal dichalcogenides such as MoS$_2$, which require complex and unscalable electrochemical setups for intercalating individual flakes,[8,12,35] our approach allows for the parallel intercalation of multiple flakes without the need for external circuitry or specific flake thickness.

As a result of the molecular insertion, the interlayer distance increases[4] (**Figure 1b**). We test three different experimental conditions for the intercalation, by (i) submerging the flakes in 2 mL of the pure AM molecules (which is liquid at room temperature) for 10 minutes; (ii) submerging the flakes in 2 mL of the pure AM molecules for 30 minutes; (iii) submerging the flakes in a 6 mL 1:2 (v/v) mixture of AM and ACN.

Experimentally, we study the enlargement of the interlayer distance by comparing the X-ray diffraction patterns of the intercalates and the pristine crystals (**Figure 1c**). We highlight that the XRD patterns were recorded for randomly distributed flakes exfoliated using an adhesive tape and transferred on a Si/SiO$_2$ substrate. The pristine as-exfoliated flakes present the lowest reflection peak at 14.67°. Using Bragg's law, we can extract the interlayer distance of each sample through the X-ray reflection's peak positions, which in the case of the pristine sample we calculate to be 6.03 Å, in good agreement with reports in the literature.[6,36,37] We note that the XRD patterns measured for pristine 2H TaS$_2$ exfoliated flakes on Si/SiO$_2$ and for a bulk 2H TaS$_2$ crystal display identical peaks (See Supporting Information Figure S1), indicating that the mechanical exfoliation and transfer on the substrate does not compromise the structural integrity of 2H-TaS$_2$.

Upon intercalation, we observe the rise of a new set of peaks, with the lowest reflection peak at 8.7° (Figure 1c). This translates into an increased interlayer distance of 10.3 Å, corresponding to an increase Δx = 4.3 Å compared to the pristine interlayer distance, hence confirming the successful intercalation process. Considering that the van der Waals gap accounts for approximately 3 Å in pristine TaS$_2$, the molecules occupy an interlayer spacing of approximately 7 Å, which corresponds to the length of fully elongated AM. Therefore, we conclude that the molecules are arranged perpendicularly to the basal plane of TaS$_2$ (Figure 1b), fully occupying the van der Waals gap.

Figure 1c shows that the new set of peaks is modified by the intercalation conditions. When the flakes are submerged in pure AM molecules for 10 minutes, the new set of peaks coexists with the pristine peak reflections. The presence of two families of peaks indicates that the intercalation is not complete and there are flakes/regions which have not been fully intercalated. The new phase presents a high crystalline quality, indicated by the width of the associated peaks, which are comparable to the reflections of the pristine phase. The pristine peaks disappear if the flakes are left for 30 min in pure AM, indicating total intercalation. However, in these conditions, the intercalated phase now shows broader peaks (10 min AM intercalation full width half maximum (FWHM)= 0.45°; 30 min AM intercalation FWHM= 1.71°), revealing lower crystallinity.

When the intercalation takes place in the AM:ACN mixture for 30 minutes, the set of pristine peaks also disappears, but in this case the peaks corresponding to the intercalated phase are narrower (FWHM= 0.76) than in the case of pure AM (FWHM= 1.71°). This demonstrates that modifying the chemical environment of the $TaS_2$ flakes during chemical intercalation yields a complete intercalation with improved crystallinity. We ascribe this improvement to two concomitant factors associated with the presence of ACN: (i) using a diluted AM intercalating medium slows the reaction kinetics, thereby reducing structural disturbance; (ii) co-intercalated ACN molecules may act as lubricants, facilitating better diffusion for AM in the van der Waals gap, and preventing oversaturation of the material with AM.

We highlight that for the intercalation in AM:ACN 1:2, we cannot rule out the presence of traces of ACN molecules co-intercalated with AM. Therefore, we refer to these samples as $(AM)_xTaS_2(ACN)_y$.

To better understand how the two intercalating environments (AM / AM:ACN 1:2) affect the exfoliated flakes, we employed optical microscopy. We recorded a 30-minute-long video of the intercalation for each of the chemical environments (Supporting video 1 and 2). It is possible to follow the progress of the intercalation in both cases through the evolution of the color hue of the TaS$_2$ flakes, an effect caused by the increasing thickness related to the insertion of molecules in the vdW gaps. In the case of the intercalation with the pure AM, the hue change is accompanied by a change in the flake morphology. While the flakes are initially flat and even, they develop a granular texture as the molecular intercalation progresses (**Figure 2a**). In contrast, in the AM:ACN 1:2 medium the change in morphology is much less evident (**Figure 2b**) as the flakes do not acquire a granular texture, yet the hue changes related to the increasing thickness are still clearly visible.

To better characterize the changes in morphologies occurring during the intercalation, we imaged a selected flake before and after intercalation by atomic force microscopy (AFM). For this experiment, we focused on the $(AM)_xTaS_2(ACN)_y$ samples, which provide a higher sample quality. The first panel of **Figure 3a** shows the morphology of the pristine 2H-TaS$_2$ flake, which has a thickness of approximately 50 nm. The second panel shows the same flake after a 30-minute-long intercalation in AM:ACN. The morphology of the intercalated flake remains relatively flat and even. Comparing the flake profiles measured before and after intercalation, we observe how the intercalated flake is now approximately 87 nm in height, showing a 70% increase in comparison to the pristine thickness (**Figure 3b**). This value is in good agreement with the increased interlayer distance measured by XRD. We note that flakes intercalated with AM without ACN show a rougher surface than those intercalated with AM:ACN (Supporting Information Figure S2), confirming the more severe structural damage introduced without ACN.

In addition to the AFM experiments we performed micro-Raman spectroscopy measurements on the same flake before and after intercalation with AM:ACN (**Figure 3c**). The pristine Raman spectrum presents the characteristic peaks of TaS$_2$[38,39] that correspond to the following lattice vibrations: 2-phonon mode (~180 cm$^{-1}$), E$_{2g}$ (~ 288 cm$^{-1}$), and A$_{1g}$ (~ 400 cm$^{-1}$). These frequencies match the values found in the literature for each peak. The two-phonon mode arises from second-order scattering of acoustic modes, while the A$_{1g}$ mode corresponds to the out-of-plane atomic displacements and the E$_{2g}$ originates from the in-plane atomic displacements (**Figure 3d**).[40]

Upon intercalation, the peak corresponding to the A$_{1g}$ mode increases in intensity and shifts towards lower wavenumber. Our observations match the data reported in the literature regarding the effects of intercalation on the Raman spectra of 2H TaS$_2$[41] and TaSe$_2$[42] intercalated with EDA and 2H-NbS$_2$ intercalated with pyridine, aniline, and picoline[43]. These reports attributed the change to a modification of long-range interaction in the superlattice introduced by the presence of highly polarizable organic molecules between host layers. Moreover, we note that the shift observed in our experiment in the out-of-plane vibration (A$_{1g}$ mode) matches the trend observed for other layered materials under pressure. In particular, the out-of-plane mode shifts to lower wavelengths in our experiment where the interlayer distance is increased, whereas it shifts to higher wavelengths in pressure experiments[44], where the interlayer distance is decreased. We highlight that the dramatic change in the Raman spectra can be used to determine whether a specific 2H-TaS$_2$ flake is successfully intercalated.

Finally, we study how the different intercalation conditions affect the electrical transport properties of the TaS$_2$ flakes. For the fabrication of the devices, we used a dry transfer technique (**Figure 4a**). After exfoliation on a polydimethylsiloxane (PDMS) block, the desired flake is transferred onto prepatterned Ti/Au contacts on a Si/SiO$_2$ substrate using an optical microscope equipped with a micromanipulator. Then, the temperature dependence of the resistance R(T) is measured in a 4-probe configuration for the pristine TaS$_2$ and for the same flake after the 30-min intercalation in either pure AM or in the AM:ACN mixture.

**Figure 4b** displays the R(T) measured in the same device before and after intercalation with pure AM for 30 minutes. The pristine flake is characterized by a metallic trend in the R(T). A change in the slope in the R(T) is observed in the vicinity of 80 K, caused by a CDW transition.[20] This anomaly can be observed in the derivative of the R(T) (inset in Figure 4b), which shows a clear dip close to the CDW temperature. The observation of this transition, which is very sensitive to both charge transfer and structural modifications, confirms that the dry transfer technique used to stamp the flake does not introduce important structural changes in the material. At low temperatures, no superconductive transition is observed in the experimentally accessible temperature range for the pristine flake (T$_{min}$ = 1.9 K, see **Figure 4c**). We note that this flake has a thickness of approximately 15 nm, so it behaves as a bulk crystal, characterized by a T$_c$ below 1 K[23] (not accessible in our experimental setup).

Once intercalated in pure AM, the room-temperature resistance of the flake increases in comparison with the pristine values. The increased resistance can likely be explained considering the structural damage introduced in the flake by the intercalation of the AM (see Figure 2). Nevertheless, as the temperature is lowered, the metallic R(T) is preserved. We highlight that the derivative of the R(T) in this case is featureless, indicating the suppression of the CDW (Inset in Figure 4b), reported for TaS$_2$ monolayers,[24] providing a first indication that the bulk intercalated TaS$_2$ acquires monolayer behavior.

Figure 4c shows the low temperature data normalized to the resistance at 10 K (normal state). The pristine device R(T) in this range is completely flat whereas the AM-intercalated device presents a superconductive transition with an onset above 3 K with a critical temperature ($T_c$) of 2.67 K. We define the $T_c$ as the temperature at which the device has dropped 50% of the normalized resistance value in the normal state. This data indicates that the superconductivity of the intercalated $TaS_2$ is strongly enhanced as compared to the pristine flake. However, despite presenting a superconductive transition, the zero-resistive state is not reached (inset of Figure 4b). This behavior is similar to what was previously observed in intercalated $MoS_2$ flakes,[12] and attributed to an inhomogeneous intercalation. In the case of $TaS_2$ intercalated in pure AM, the absence of a zero-resistance state can be explained based on the structural damage introduced by the intercalation (see Figure 2), which may locally inhibit the superconducting phenomenon, interrupting the superconducting path.

**Figure 4d** contains the R(T) of another $TaS_2$ device, measured before and after intercalation in AM:ACN. Similarly to the case of intercalation in pure AM, the CDW phase is suppressed after intercalation and both the pristine and the intercalated flake present a metallic behavior as the temperature drops. Notably, the room temperature resistance of the $(AM)_x TaS_2 (ACN)_y$ flake is lower in the intercalated than in the pristine state, unlike the case of pure AM. The lower resistance can be related to charge carrier doping introduced by the presence of amine groups, which are often used as n-type dopants for 2D materials.[45–48] Moreover, the lower resistance confirms that the AM:ACN treatment introduces a lower degree of structural damage as compared to the pure AM case.

The low temperature evolution of the resistance is also significantly different. While the pristine device does not show any superconducting transition in the experimentally available temperature range, $(AM)_x TaS_2 (ACN)_y$ presents a superconductive transition with a $T_c = 2.91$ K, slightly higher than the $T_c$ of the pure AM case, 2.67 K. Moreover, the transition is steeper, and most importantly, (AM)$_x$TaS$_2$(ACN)$_y$ does reach a zero-resistance state (see the inset included in **Figure 4e**). This can be understood considering that the gentler AM:ACN intercalation process introduces fewer structural defects, therefore enabling the spreading of the superconducting state to the whole TaS$_2$ flake. We note that the enhanced T$_c$ reached in our intercalated flakes is similar to that of pristine monolayer TaS$_2$.[24] This monolayer behavior recorded in tens-of-nm thick crystals can be ascribed to the molecular intercalation, which separates the individual planes and minimizes the interlayer interaction, resulting in bulk flakes behaving as a stack of monolayers.[8]

Moreover, it should be noted that the charge carrier doping effect introduced by the AM intercalation also contributes to the enhancement of superconductivity. Superconducting transitions are highly sensitive to doping levels, as evidenced by the characteristic dome-like relation between Tc and charge carrier concentration observed for instance in cuprate superconductors.[49] More recently, similar phase diagrams have been observed in different systems through doping via electrostatic gating.[50,51] In our experiments, charge doping and increased interlayer separation synergistically enhance the measured Tc, although it is challenging to disentangle the individual contributions of these two effects.

We highlight that intercalation has a key advantage with respect to the isolation of monolayers to obtain TaS$_2$ devices with a superconductivity onset higher than 3 K. Individual monolayers are air sensitive, and therefore they require complex fabrication and careful encapsulation in a controlled environment.[24] On the contrary, our intercalated flakes, while suitable for integration in complex device architectures and displaying a T$_c$ similar to monolayers, are processed in air and do not require any encapsulation.

4. **Conclusions**

In summary, our study presents a straightforward technique for the in-situ intercalation of 2H-TaS$_2$ flakes to produce superconductive devices. The comparison between two intercalation environments reveals that reducing the concentration of the intercalating agent (AM) using a solvent (ACN) results in a gentler molecular insertion. In turn, this leads to a lower degree of structural disorder and damage, as confirmed by XRD and optical microscopy. Notably, the intercalation conditions also affect the electrical transport measurements. While the intercalation in pure AM causes an incomplete superconducting transition, the AM:ACN treatment leads to a fully developed zero-resistance state. Our method offers a reliable, reproducible, and rapid approach for obtaining TaS$_2$-based superconductive devices, opening the way to the integration of superconducting intercalation flakes into functional quantum devices. Moreover, our data indicate that the intercalation of 2H-TaS$_2$ can also be explored for the formulation of functional superconducting inks.[52]

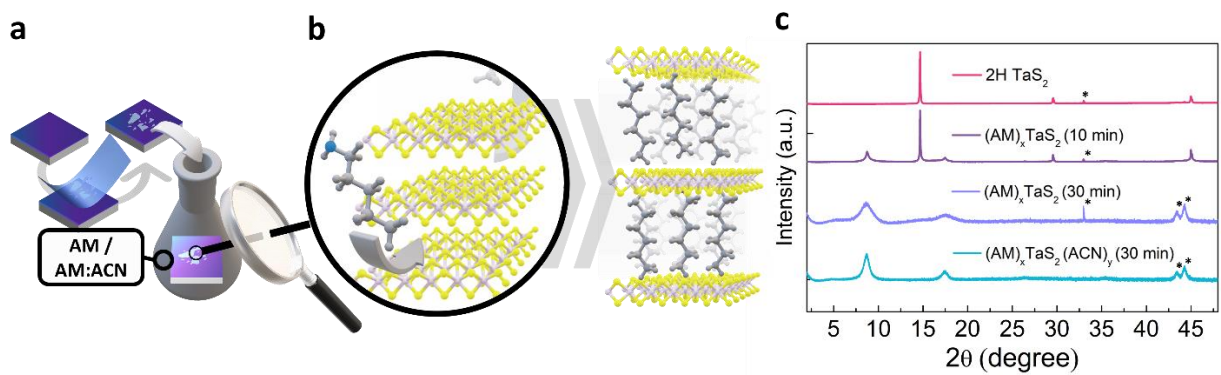

**Figure 1.** a) Exfoliation of 2H-TaS$_2$ flakes on the surface of SiO$_2$/Si substrate followed by submersion in amylamine (AM) or amylamine:acetonitrile (AM:ACN) mixture. b) Sketch of the insertion of AM molecules in the interlayer gaps of the hosting 2H-TaS$_2$ flake. c) X-ray diffraction patterns of pristine flakes, (AM)$_x$TaS$_2$ flakes obtained after dipping the sample for 10 and 30 minutes in an AM solution, and (AM)$_x$TaS$_2$(ACN)$_y$ flakes obtained after dipping the sample for 10 and 30 minutes in an AM solution. Spurious peaks originating from the substrate are labeled with a star. The patterns of (AM)$_x$TaS$_2$ and (AM)$_x$TaS$_2$(ACN)$_y$ were multiplied by a factor of 20 to make their intensity comparable to that of the pristine flakes.

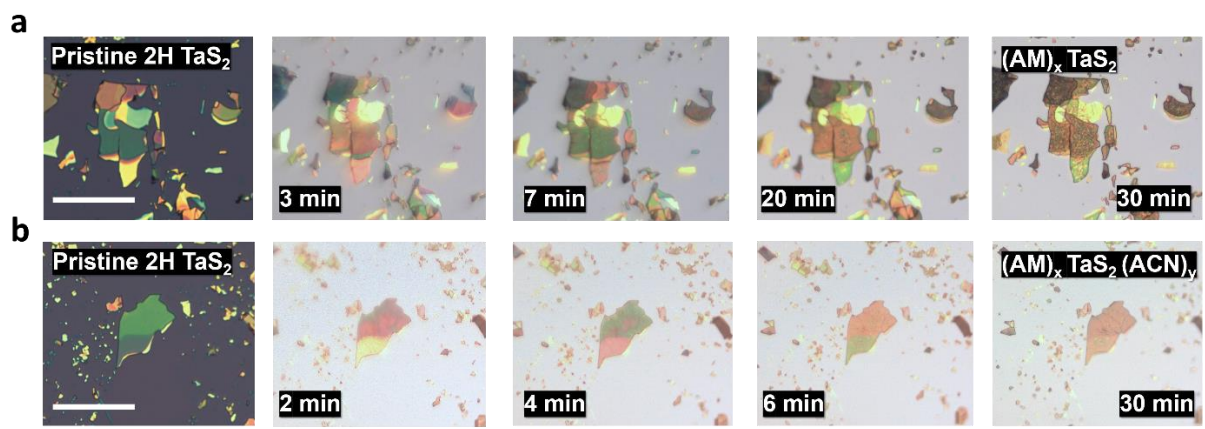

**Figure 2.** a) Optical imaging of the intercalation of 2H-TaS$_2$ with pure AM for 30 minutes. b) Optical imaging of the intercalation of 2H-TaS$_2$ with AM:ACN 1:2 for 30 minutes. Scale bar: 20 µm.

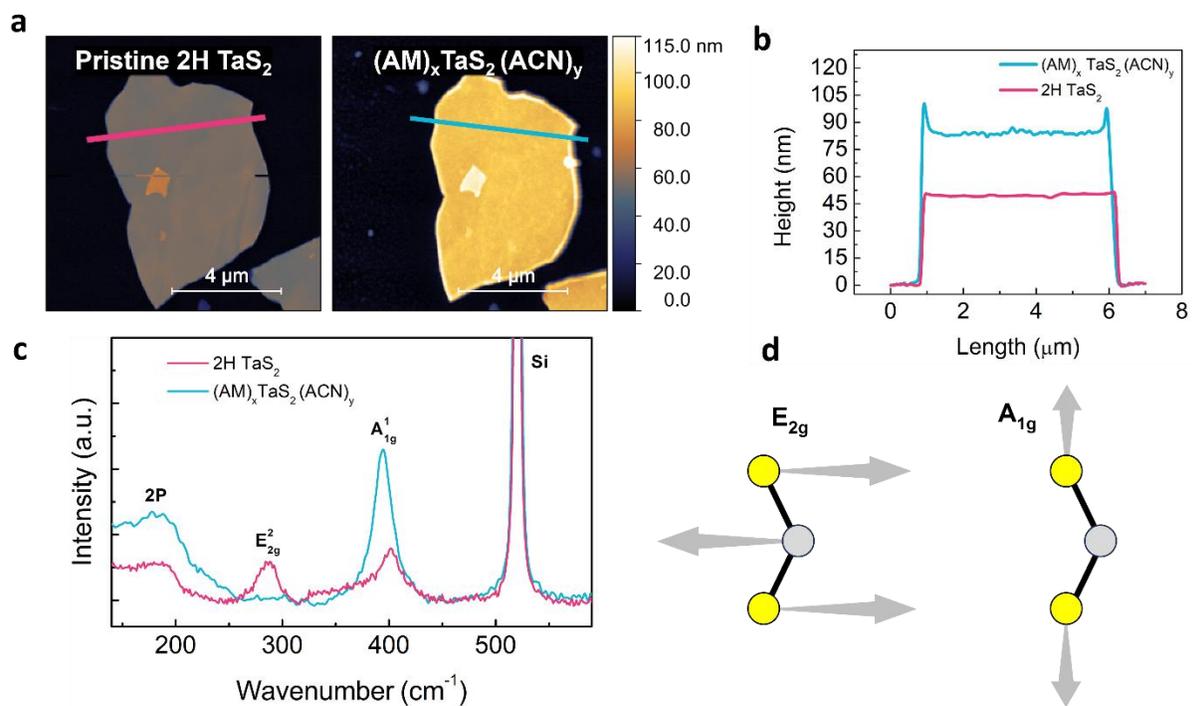

**Figure 3.** a) Atomic force microscopy image of a 2H-TaS$_2$ flake before and after intercalation with mixed medium (AM:ACN). b) Height profile of the pristine and intercalated flake. c) Raman spectra of the pristine and intercalated flake. d) Atomic displacement vibrations corresponding to the Raman modes E$_{2g}$ and A$_{1g}$.

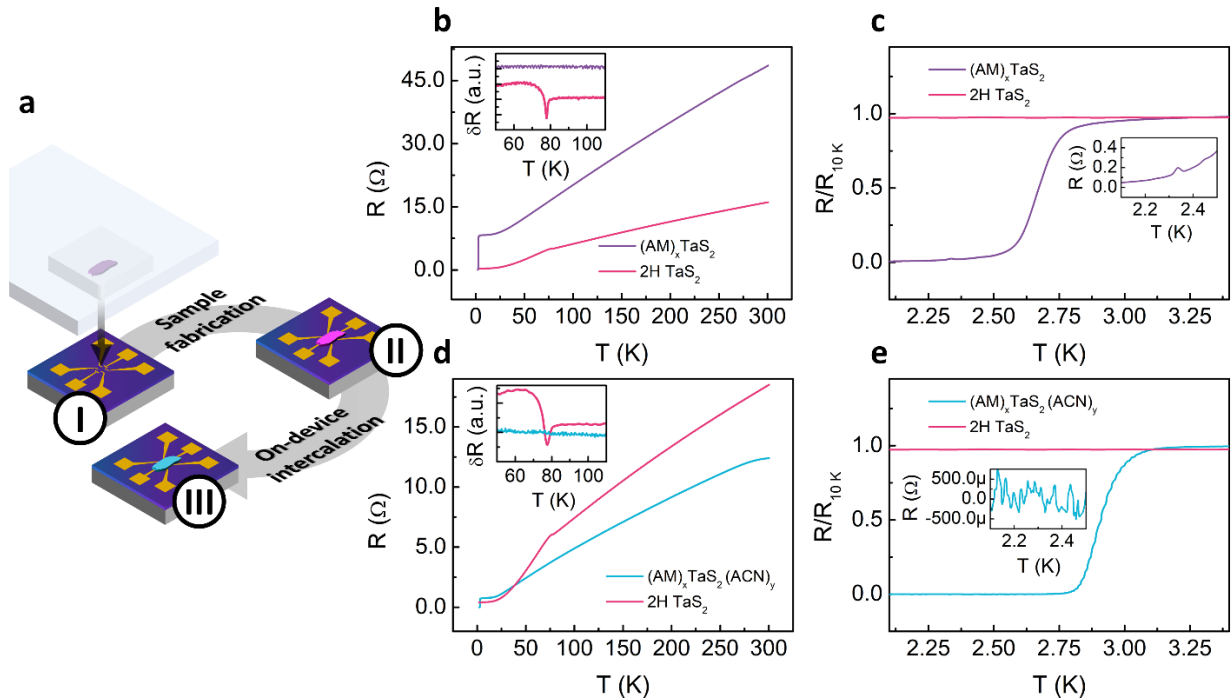

**Figure 4.** a) Sample preparation process involves (I) dry stamping of the desired flake onto the prepatterned Si/SiO$_2$ substrates, (II) low temperature measurement of the pristine device, and (III) intercalation & low temperature measurement of the intercalate. b) Temperature dependence of the resistance R(T) of a TaS$_2$ flake from 1.9 K to 300 K before and after 30 min in pure AM. Inset: derivative of R(T) in the range of 50 to 110 K. c) R(T) normalized to the resistance at 10 K of the same flake in the range from 1.9 K to 3.5 K, including an inset with the R(T) (raw data) in the range of 2.1 to 2.5 K. d) R(T) of a TaS$_2$ flake from 1.9 K to 300 K before and after 30 min in AM:ACN 1:2. Inset: derivative of R(T) in the range of 50 to 110 K. e) R(T) normalized to the resistance at 10 K of the same flake in the range from 1.9 K to 3.5 K, including an inset with the R(T) (raw data) in the range of 2.1 to 2.5 K.

## ASSOCIATED CONTENT

**Supporting Information**

Video 1 shows the intercalation of 2H-TaS$_2$ flakes in pure amylamine (AM).

Video 2 shows the intercalation of 2H-TaS$_2$ flakes in the amylamine:acetonitrile (AM:ACN) mixture.

## AUTHOR INFORMATION

**Corresponding Author**

*Marco Gobbi − IKERBASQUE, Basque Foundation for Science, 48013 Bilbao, Spain; Centro de Física de Materiales (CFM-MPC) Centro Mixto CSIC-UPV/EHU, 20018 Donostia-San Sebastián 20018, Spain; https://orcid.org/0000-0002-4034-724X; Email: marco.gobbi@ehu.eus

**Authors**

Jose M. Pereira – CIC nanoGUNE BRTA, 20018 Donostia-San Sebastian, Spain, https://orcid.org/0000-0002-5179-2349

Daniel Tezze – CIC nanoGUNE BRTA, 20018 Donostia-San Sebastian, Spain, https://orcid.org/0000-0002-9153-6219

Beatriz Martín-García − IKERBASQUE, Basque Foundation for Science, 48013 Bilbao,; CIC nanoGUNE BRTA, 20018 Donostia-San Sebastian, Spain, https://orcid.org/0000-0001-7065-856X

Fèlix Casanova − IKERBASQUE, Basque Foundation for Science, 48013 Bilbao; CIC nanoGUNE BRTA, 20018 Donostia-San Sebastian, Spain, https://orcid.org/0000-0003-0316-2163

Maider Ormaza − Departamento de Polímeros y Materiales Avanzados: Física, Química y Tecnología (UPV-EHU), 20018 San Sebastián, Spain, https://orcid.org/0000-0002-7278-5308

Luis E. Hueso − IKERBASQUE, Basque Foundation for Science, 48013 Bilbao, Spain; CIC nanoGUNE BRTA, 20018 Donostia-San Sebastian, Spain, https://orcid.org/0000-0002-7918-8047

**Notes**

The authors declare no competing financial interest.

**Funding Sources**

ACKNOWLEDGMENT

This work was supported under Projects PID2021-128004NB-C21 and PID2021-122511OB-I00 funded by Spanish MCIN/AEI/10.13039/501100011033 and by ERDF A way of making Europe; and under the María de Maeztu Units of Excellence Programme (Grant CEX2020-001038-M). This work was also supported by the FLAG-ERA grant MULTISPIN, via the Spanish MCIN/AEI with grant number PCI2021-122038-2A. B. M.-G. and M. G. acknowledge support from the "Ramón y Cajal" Programme by the Spanish MCIN/AEI (grant no. RYC2021-034836-I and RYC2021-031705-I).

TOC

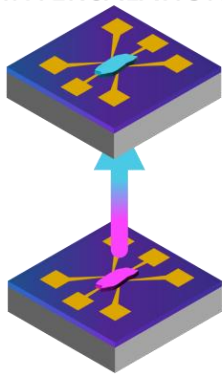 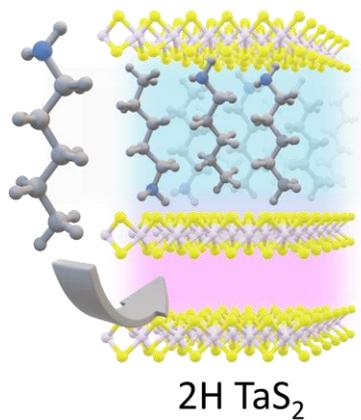 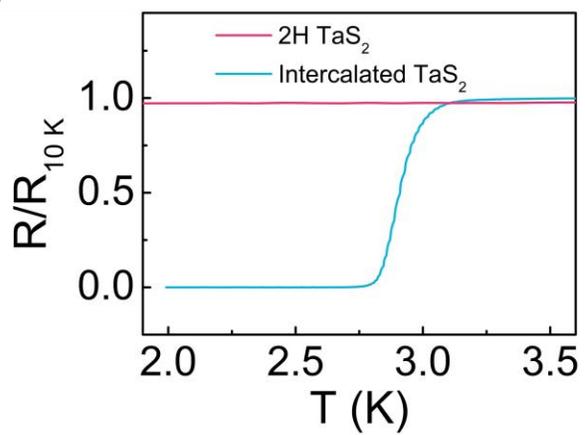